\documentclass[12pt,a4paper,fleqn]{article}
\usepackage{tikz}
\usepackage{caption}
\usepackage{lscape}

\usepackage{fullpage}
\usepackage{amsmath}
\usepackage{amsthm}
\usepackage{amssymb}
\usepackage{amscd}
\usepackage{plain}
\usepackage{graphicx}
\usepackage{t1enc,epsfig}
\usepackage[mathscr]{eucal}
\usepackage{epstopdf}
\usepackage[german,english]{babel}
\usepackage{stmaryrd}

\usepackage{xcolor}

\usepackage{wrapfig}
\usepackage{float}

\numberwithin{equation}{section}

\newenvironment{proof1}{{\bf Proof. }}{\hfill$\rule{1ex}{1ex}$\par\medskip}

\newtheorem*{lema}{Lemma 1}
\newtheorem*{lemb}{Lemma 2}
\newtheorem*{lemc}{Lemma 3}
\newtheorem*{lemd}{Lemma 4}
\newtheorem*{leme}{Lemma 5}

\newtheorem*{thma}{Theorem 1}
\newtheorem*{thmb}{Theorem 2}
\newtheorem*{thmc}{Theorem 3}

\newtheorem*{ra}{Remark 1}
\newtheorem*{rb}{Remark 2}
\newtheorem*{rc}{Remark 3}
\newtheorem*{rdd}{Remark 4}

\begin{plain}

\def\Z  {{\mathchar"0\hexnumber@\msbfam5A}} 
\def\N  {{\mathchar"0\hexnumber@\msbfam4E}} 
\def\A  {{\mathchar"0\hexnumber@\msbfam41}} 
\def\B  {{\mathchar"0\hexnumber@\msbfam42}} 
\def\C  {{\mathchar"0\hexnumber@\msbfam43}} 
\def\D  {{\mathchar"0\hexnumber@\msbfam44}} 

\def\R  {{\mathchar"0\hexnumber@\msbfam52}} 
\def\be {{\mathchar"0\hexnumber@\mibfam65}}
\def\bo {{\mathchar"0\hexnumber@\mibfam6F}}
\def\bu {{\mathchar"0\hexnumber@\mibfam75}}
\def\bv {{\mathchar"0\hexnumber@\mibfam76}}
\def\bw {{\mathchar"0\hexnumber@\mibfam77}}
\def\bx {{\mathchar"0\hexnumber@\mibfam78}}
\def\by {{\mathchar"0\hexnumber@\mibfam79}}
\def\bz {{\mathchar"0\hexnumber@\mibfam7A}}
\def\e {t}

\def\diy{\displaystyle}
\def\dfrac{\diy\frac}

\def\strike#1{\leavevmode\vbox{\hbox{#1}\kern-0.5ex\hrule height 0.4pt\kern0.5ex}}

\input colordvi
\end{plain}

\begin{document}
%%%%%%%%%%%%%%%%%%%%%%%%%%%%%%%%%%%%%%%
\newcommand{\blema}{\begin{lema}}
\newcommand{\elema}{\end{lema}}
\newcommand{\blemb}{\begin{lemb}}
\newcommand{\elemb}{\end{lemb}}
\newcommand{\blemc}{\begin{lemc}}
\newcommand{\elemc}{\end{lemc}}
\newcommand{\blemd}{\begin{lemd}}
\newcommand{\elemd}{\end{lemd}}
\newcommand{\bleme}{\begin{leme}}
\newcommand{\eleme}{\end{leme}}

\newcommand{\bthma}{\begin{thma}}
\newcommand{\ethma}{\end{thma}}
\newcommand{\bthmb}{\begin{thmb}}
\newcommand{\ethmb}{\end{thmb}}
\newcommand{\bthmc}{\begin{thmc}}
\newcommand{\ethmc}{\end{thmc}}

\newcommand{\bra}{\begin{ra}}
\newcommand{\era}{\end{ra}}
\newcommand{\brb}{\begin{rb}}
\newcommand{\erb}{\end{rb}}
\newcommand{\brc}{\begin{rc}}
\newcommand{\erc}{\end{rc}}
\newcommand{\brdd}{\begin{rdd}}
\newcommand{\erdd}{\end{rdd}}
%%%%%%%%%%%%%%%%%%%%%%%%%%%%%%%%%%%%

\def\Z{{\mathbb Z}} 
\def\A{{\mathbb A}} 
\def\R{{\mathbb R}}
\def\N{{\mathbb N}}  
\def\e {t}
\def\bx{{\mathbf x}}
\def\bv{{\mathbf v}}
\def\by{{\mathbf y}}
\def\bz {{\mathbf z}}
\def\bx{{\mathbf x}}
\def\bo{{\mathbf o}}
\def\be {{\mathbf e}}
\def\bw {{\mathbf w}}
\def\bma{\begin{matrix}} \def\ema{\end{matrix}}

\title{\bf A Reverse Hard-Core Model in $\A_2$: \\
an Application of the Pirogov-Sinai Theory}

\author{\bf A. Mazel$^1$, I. Stuhl$^2$, Y. Suhov$^{2,3}$}

\date{}
\footnotetext{2010 {\em Mathematics Subject Classification:\; primary 60G60, 82B20, 82B26}}
\footnotetext{{\em Key words and phrases:} reverse hard-core model, particle activity, extreme Gibbs measure, ground state, Pirogov-Sinai theory, contour, Peierls estimate, polymer expansion

\noindent
$^1$ AMC Health, New York, NY, USA;\;\;
$^2$ Math Dept, Penn State University, PA, USA;\;\;
$^3$ DPMMS, University of Cambridge and St John's College, Cambridge, UK}

\maketitle

\begin{abstract} 
In this paper we use the Pirogov--Sinai theory to analyze a class of particle models of Statistical Mechanics on the unit triangular lattice $\A_2$. The models are specified by two parameters: the particle activity $u\in(0,1)$ and a positive L\"oschian number $\e$ interpreted as a maximal squared clearing radius. Admissible configurations are those in which every empty lattice site has an occupied site within squared distance at most $\e$, thereby forbidding empty disks of radius exceeding or equal $\e$. The Hamiltonian favors configurations with as few occupied sites as possible, while the admissibility condition enforces a positive density of particles.

We prove that the periodic ground states of the model, i.e., the configurations achieving an optimal balance of two tendencies, are the ones whose occupied sites form triangular sublattices of explicitly determined squared side-length $d_*(\e)$. Furthermore, for sufficiently small values of the activity parameter $u$, some (but not necessarily all) periodic ground states generate extreme Gibbs (DLR) measures. We describe the resulting phase structure and characterize the pure phases associated with stable ground states. 
\end{abstract}

\section{Introduction. A reverse hard-core model}\label{Intro}

We study statistical-mechanical models 
on a unit triangular lattice $\A_2=\{\bx=m\be_1+n\be_2:\;m,n\in\Z\}\subset\R^2$
where $\be_1=(1, 0)$ and $\be_2= (1/2, {\sqrt 3}/2)$ in Cartesian coordinates. 
In the rest of the paper we utilize the $\A_2$-coordinates such that 
$\bx=(m,n)$. 
The corresponding $\A_2$-norm is $N(\bx)=N(m,n) :=m^2+n^2+mn$; it yields
the squared Euclidean distance $|\bx -\bo|^2$ between $\bx$ and the origin $\bo$.

Given $\e >0$ and $\bz\in\R^2$, define  
$B_\e(\bz)=\{\by \in \R^2:\; |\bz-\by|^2 \le \e\}$, a closed $\R^2$-disk of the squared 
radius $\e$ centered at $\bz$. 

The object of study in this paper is a class of models on $\A_2$ we call {\it reverse 
hard-core} or {\it bounded-void} models. Formally, 
a configuration $\phi \in \{0,1\}^{\A_2}$ is called {\it $\e$-admissible} if for any $\bz \in \A_2$ there exists  $\bx \in B_{\e}(\bz)\cap\A_2$
such that $\phi(\bx) = 1$. In other words, one does not allow void disks 
$B_\e(\bz)$ for which $\phi$ takes a constant value $0$. 
Given a $\e$-admissible 
configuration $\phi$, we say a site $\bx\in\A_2$ is {\it occupied} (and write 
$\bx\in\phi$), if $\phi(\bx)=1$, and say $\bx$ is
{\it vacant} or {\it empty} otherwise. Correspondingly, $\e$ stands for 
``emp{\bf t}y''. The set 
$\Phi(t)$ of $\e$-admissible configurations is a measurable subset of
$\{0,1\}^{\A_2}$, i.e., $\Phi (t)$ belongs to the sigma-algebra generated by the
cylinder subsets of $\{0,1\}^{\A_2}$.
The above definition of a $\e$-admissible configuration remains in force if we use a subset 
of lattice $\A_2$.

Fix a particle activity $u\in (0,1)$ and set
$$
H(\phi) = -\log(u) \sum_{\bx \in \A_2} \phi(\bx),\quad\phi\in\Phi (t).\eqno(1)
$$

The formal Hamiltonian $H$ defines a lattice model of statistical mechanics
on the set $\Phi(t)$. In this model the value $u$ favors as few occupied sites as
possible while the admissibility condition 
forces the $\R^2$-density of occupied sites to be at least $1\big/|B_\e(\cdot)|$, where $|\cdot |$ denotes the $\R^2$-area. In what follows, the L\"oschian number $\e$, for most of
the time, is taken to be fixed, and we write $\Phi$ instead of $\Phi (t)$.

In a sense, the class of models with Hamiltonians as in (1) is dual of the class of 
traditional hard-core (HC) models where 
the HC requirement reads: $|\phi(\bx)-\phi(\by)|^2 \ge d$  if 
$\phi(\bx)=\phi(\by)=1$, and activity $u>1$. An emerging question is: what are the extreme Gibbs (DLR) measures for the dual model? The key observation of this paper is that the Pirogov-Sinai 
(PS) theory covers both classes of models in a similar fashion, which allows us to use the 
approach developed in \cite{[MSS1]}, \cite{[MSS2]} and answer the above question for the dual models with
$u<1$.

Models of this type have been studied in recent papers \cite{[WBK1]}, \cite{[WBK2]}, motivated by applications to recovery of information in coding systems. Papers \cite{[WBK1]}, \cite{[WBK2]} analyzed the case of $\e  =1$.  
The present work can be considered as 
an extension of \cite{[WBK2]} to the case of an arbitrary L\"oschian number $\e$.

The main results of the current paper are Theorems 1--3; see 
Section 2. These 
theorems establish the structure of the phase diagram for the Gibbs distributions on 
the set $\Phi$, generated by  
Hamiltonian (1) for values $u\in (0,u_0)$ where $u_0\in (0,1)$ depends on $\e$.

\section{The PS theory for the reverse HC model}\label{PS}

The appproach adopted in this paper is based on
representing a configuration $\phi$ by a triangulation of $\A_2$. The triangulations 
under consideration are uniquely defined modifications of the Delaunay 
triangulation \cite{[AKL]}. All triangles in triangulations are considered to be closed.
 The key observation is that 
the area of each triangle in such a triangulation is bounded from
above and attains its maximum value if and only if $\phi = \varphi :\simeq 
\sqrt{d_*(\e )}\cdot \A_2$. Here, $d_*(\e )$ denotes the Löschian number defined by equations (6) 
and (7) 
in Section 3. 

This places us in a position to rewrite the model in terms of a convenient $m$-{\it potential}
(cf. \cite{[HS]}, \cite{[MSS1]}). Given $\e$, we set $d=d_*(\e)$ until the end of this 
section. The $m$-potential is defined on disks $B_{4d}(\cdot)$. Given $\bz \in A_2$ and a
$\e$-admissible configuration $\phi_{B_{4d}(\bz)}$ in $\A_2\cap B_{4d}(\bz)$, consider the set of 
the occupied sites $\bx\in\phi_{B_{4d}(\bz)}$ and the corresponding triangulation. Let 
$\triangle (\bz)=\triangle (\bz,\phi_{B_{4d}(\bz)})$ be the collection of triangles from 
the triangulation of the set of the occupied sites in $\phi_{B_{4d}(\bz)}$, which 
contains $\bz$. If $\bz$ is vacant, then  
$\triangle (\bz)$ contains one or two triangles. If $\bz$ is occupied, then 
$\triangle (\bz)$ contains at least three triangles.

For any $\bz \in \A_2$, let 
$C(\bz) \in \R^2$ be the corresponding Voronoi cell of $\A_2$, considered as a collection of 
points in $\R^2$. Clearly, $C(\bz)$ is a hexagon with 
the edges of length $1/\sqrt{3}$. Set
$$U\left(\bz|\phi_{B_{4d}(\bz)}\right)=\sum_{T\in\triangle (\bz)} 
{|C(\bz)\cap T| \over 2|T|},\eqno (2)$$ 
where the sum is taken over all triangles $T$ of $\phi_{B_{4d}(\bz)}$, containing $\bz$. 
Hamiltonian $H$ (see (1)) can be formally re-written as
$$H(\phi)=-\log(u)\sum_{\bz \in \A_2} U\left(\bz|\phi_{B_{4d}(\bz)})\right),\quad
\phi\in\Phi . \eqno (3)$$
In fact, the sum in the right hand side of (3) %(5 iii) 
equals a half of the number of triangles 
in $\phi$ which corresponds to the occupied sites in $\phi$. Lemma 4 in Section 4 implies that 
$$\min_{\phi, \bz\in \A_2} U\left(\bz|\phi_{B_{4d}(\bz)})\right)=\dfrac{1}{d}.$$
Furthermore, 
$$U\left(\bz|\varphi_{B_{4d}(\bz)})\right)={1 \over d}$$
for any $\bz\in\A^2$ and $\varphi \cong \sqrt{d}\cdot\A_2$. It means that 
$U(\cdot|\cdot)$ is 
an $m$-potential, and configurations  $\varphi \cong \sqrt{d}\cdot\A_2$ are the only 
corresponding {\it perfect} configurations, i.e., those providing the minimal value of 
$U(\bz|\cdot)$ for every $\bz\in \A_2$. Observe that each perfect configuration gives a
periodic ground state and vice versa.

Let $s=\max\, |T|$, where $T$ is an $\A_2$-triangle that is not a $\sqrt{d}$-equilateral one. Lemma 4 implies that $0<s < d\sqrt{3}/4$. The positive quantity
$$\tau :=\dfrac{\sqrt{3}}{4s} -\dfrac{1}{d}\eqno (4)$$%\eqno (6 iv)$$
is the {\it Peierls constant} corresponding to the $m$-potential $U(\cdot|\cdot)$. 
That is, the {\it Peierls estimate} holds true: 
$$U\left(\bz|\phi_{B_{4d}(\bz)})\right)-U\left(\bz|\varphi_{B_{4d}(\bz)})\right)
\geq\tau \eqno (5)$$ %\eqno (7 v)$$
whenever $U(\bz|\phi_{B_{4d}(\bz)})\neq U(\bz|\varphi_{B_{4d}(\bz)})$.

This observation places the model within the well-developed framework of the PS theory; see, for example, \cite{[PS]}, \cite{[Z]}, \cite{[HS]}, \cite{[MSS2]}, and \cite{[DS]}. The PS theory yields a number of results for reverse HC models analogous to the theorems established in \cite{[MSS1]}. In particular, these results describe periodic ground states (PGSs) and extreme Gibbs measures (EGMs) on $\Phi$ for Hamiltonian $H$ as in (1) and (3). %(5 iii). 
Bound (5), %(7 v), 
together
with Lemmas 1 -- 5 in Sections 3, 4, implies the following theorem.

\bthma\label{Theorem 1.} 
For any L\"oschian number $\e$, the set of the occupied sites in a  
PGS for Hamiltonian $H$ is congruent to $\sqrt{d_*(\e )}\cdot\A_2$. The PGSs are
 partitioned into equivalence classes relative to $\A_2$-shifts and $\A_2$-symmetries. 
Each equivalence class contains $2d_*(\e )$ configurations. The number of 
equivalence classes depends on the value $\e$  and is always finite. 
\ethma

According to the standard framework of the PS theory, Theorem~1 and the Peierls estimate (5) %(7 v) 
imply Theorems 2 and 3 below. We say that an EGM $\mu$ is generated by a PGS $\varphi$ if $\mu( \Theta_\varphi)=1$, where $\Theta_{\varphi}\subset  \Phi$ is formed by $\e$-admissible configurations $\phi$ such that the set of sites $\bx\in\A_2$ where $\phi(\bx) = \varphi(\bx)$ has an infinite connected component, while the set of sites $\bx\in\A_2$ where $\phi(\bx) \neq \varphi(\bx)$ has no infinite connected component.

A PGS equivalence class from Theorem 1 is called {\it stable} if each PGS from this class generates 
a distinct EGM.

\bthmb\label{Theorem 2.} 
Suppose that for a given $\e$ there exists a unique equivalence
class of PGSs. Then there exists $u_0(t)\in (0,1)$ such that for $u\in (0,u_0(\e ))$ 
each PGS generates a periodic EGM, and there exist no other EGMs.
\ethmb

\bthmc\label{Theorem 3.} 
Suppose that for a given $\e$ there exist at least two equivalence
classes of PGSs. Then there exists $u_0(t)\in (0,1)$ such that for $u\in (0,u_0(\e ))$, at least one of these classes is stable, and any EGM is generated by a PGS from a stable equivalence class.
\ethmc

According to the PS theory, the truncated correlation functions for contours
in the EGMs from Theorems 2, 3 are explicitly expressed in terms of convergent 
polymer expansions. Consequently, the truncated correlation functions decay
exponentially fast. 

\bra\label{Remark 2.} 
{\rm There is no simple criteria to decide if a given L\"oschian number $\e$ belongs to the case covered by Theorem~2 or to the case covered by Theorem~3. 

It is clear from our construction that each $\e=N(n-1,n)=3n^2-3n+1$, $n=1,2,\ldots$ belongs to Theorem~2, and the corresponding value $d_*(\e)=9n^2-3n+1$. Hence, there are infinitely many values $\e$ belonging to Theorem~2. 

Furthermore,  a direct lattice enumeration shows that for $\e=373$ and the corresponding value $d_*(\e)=1183=7\cdot 13^2$ one has two classes. They are given by $\sqrt{d_*(t)}$-equilateral $\A_2$-triangles with vertices at sites $(4,17)$ and $(8,14)$, respectively. Thus, the set of values $\e$ from Theorem~3 is not empty. Note that $t=372$ belongs to Theorem~2. A non-trivial fact is that the set of values $\e$ from Theorem~3 is infinite; cf. Remark 4 in Section 4.} 
\era

\brb\label{Remark 3.}  
{\rm We expect that
under the conditions of Theorem~3, there exists a single stable
PGS equivalence class. Indeed, the PS theory provides a constructive perturbation theory (see \cite{[Z]}) to 
identify the PGSs generating EGMs. The entire construction is invariant under $\A_2$-shifts and $\A_2$-symmetries. Consequently, each corresponding PGS generates a distinct EGM. Conversely, if the perturbation series associated with two equivalence classes yields the same free energy, it would imply existence of infinitely many essentially geometric identities at every order of the perturbation expansion. Such a hidden symmetry in $\A_2$ appears highly unlikely. Note that our model does not have an additional parameter (like an external field) which can be adjusted to equalize two asymmetric perturbative free energies.}
\erb

\brc\label{Remark 4.} 
{\rm The above results remain valid if one adds an additional HC interaction 
that still allows particles to be at distance $\geq d_*(\e )$. One such example is a HC model from 
\cite{[WBK2]}, where any two
occupied sites are at the squared distance $>1$ from each other and an additional 
property holds, that all configurations are saturated, i.e., it is impossible to add another 
particle to the configuration without breaking the HC condition. In this case the saturation property is equivalent to the current admissibility requirement that
a configuration does not have a vacant disk $B_1 (\bz)$. Additionally, both the standard 
HC model and the reverse HC model come together. Consequently, the approach from \cite{[MSS2]}, \cite{[MSS1]}
and the approach from the current paper cover the cases  
 $u\gg 1$ and $u\ll 1$, respectively. However, the structure of the phase
diagram for intermediate values $u$ remains an open question.}
\erc

\section{Strongly $\e$-admissible triangles}\label{strongly_t_adm}

The identification of the ground states plays a central role in the framework of the PS theory. To this end, we say that an $\A_2$-triangle $T$ is {\it $\e$-admissible} if it can be found in the Delaunay triangulation of the set of occupied sites for at least one $\phi\in \Phi$. This definition establishes the class of local geometric objects compatible with the constraints imposed by the model.

The concept of $\e$-admissibility is essential for the subsequent classification of local ground-state patterns and for the derivation of contour representations within the PS formalism. Indeed, not every geometrically possible triangle can occur in a physically realizable configuration. The admissibility condition imposes a nontrivial restriction on the local structure of the triangulation. The property formulated in Lemma 1 below provides a necessary condition for the $\e$-admissibility of an $\A_2$-triangle.

\blema\label{Lemma 1.} 
The circumradius of any $\e$-admissible 
triangle is smaller than $\sqrt{\e}+1\big/\sqrt{3}$.
\elema

\begin{proof1}
Consider 
a unit $\A_2$-triangle $W$ containing the circumcenter $\bz$ of a Delaunay triangle $T$ 
in a configuration $\phi\in\Phi$. The distance from $\bz$ to at least one of the vertices 
of $W$ is $\leq1/\sqrt{3}$; thus, this vertex lies in a closed disk $B_r(\bz)$ of squared radius $r\leq 1/3$ centered at $\bz$. If the circumradius of $T$ is larger than $\sqrt{\e} + {1/\sqrt{3}}$, then $B_r(\bz)$ is at distance $>\sqrt{\e}$ from the closed exterior of the circumcircle, hence from any occupied site. This is impossible as it breaks $\e$-admissibility of $\phi$.
\end{proof1}

An $\A_2$-triangle is called {\it strongly $\e$-admissible} if all 
$\A_2$-sites belonging to its $\R^2$-closure are at squared distance $\leq\e$ from the union of its vertices. A strongly $\e$-admissible triangle is $\e$-admissible because one can extend it to a sublattice of $\A_2$ and such a configuration is $\e$-admissible by construction (see Figure~1). The assertion of Lemma 2 gives a sufficient condition of strong $\e$-admissibility. 

\begin{figure}[H]
\begin{center}
\includegraphics[scale=0.6]{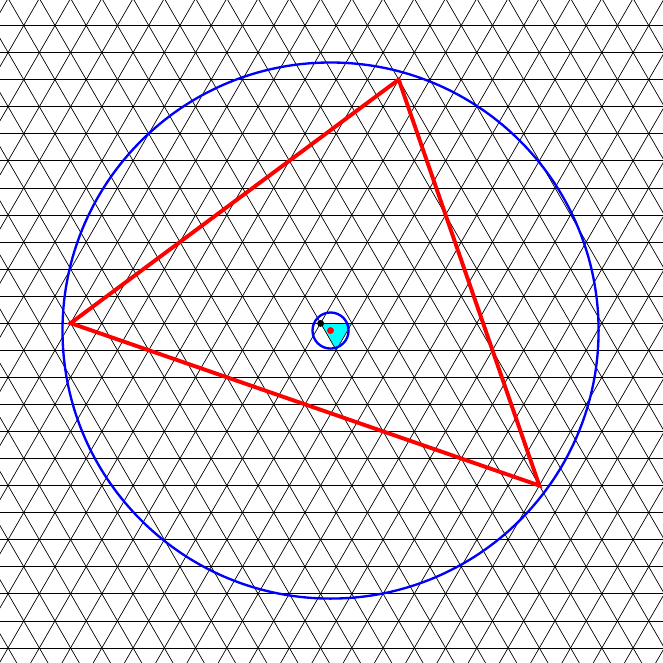}
\caption{A $\e$-admissible triangle with circumcenter $\bz$ and the corresponding disk $B_r(\bz)$.} 
\end{center}
\end{figure}

\blemb\label{Lemma 2.} 
If the squared circumradius of an $\A_2$-triangle is $\leq\e$, then this triangle 
is strongly $\e$-admissible.
\elemb

\begin{proof1}
By assumption,  the circumcenter of the triangle and all points inside triangle's $\R^2$-closure belong to the union of closed disks of squared radius at most $\e$ centered 
at triangle's vertices (see Figure~2).
\end{proof1}

\begin{figure}[h]
\begin{center}
\includegraphics[scale=0.4]{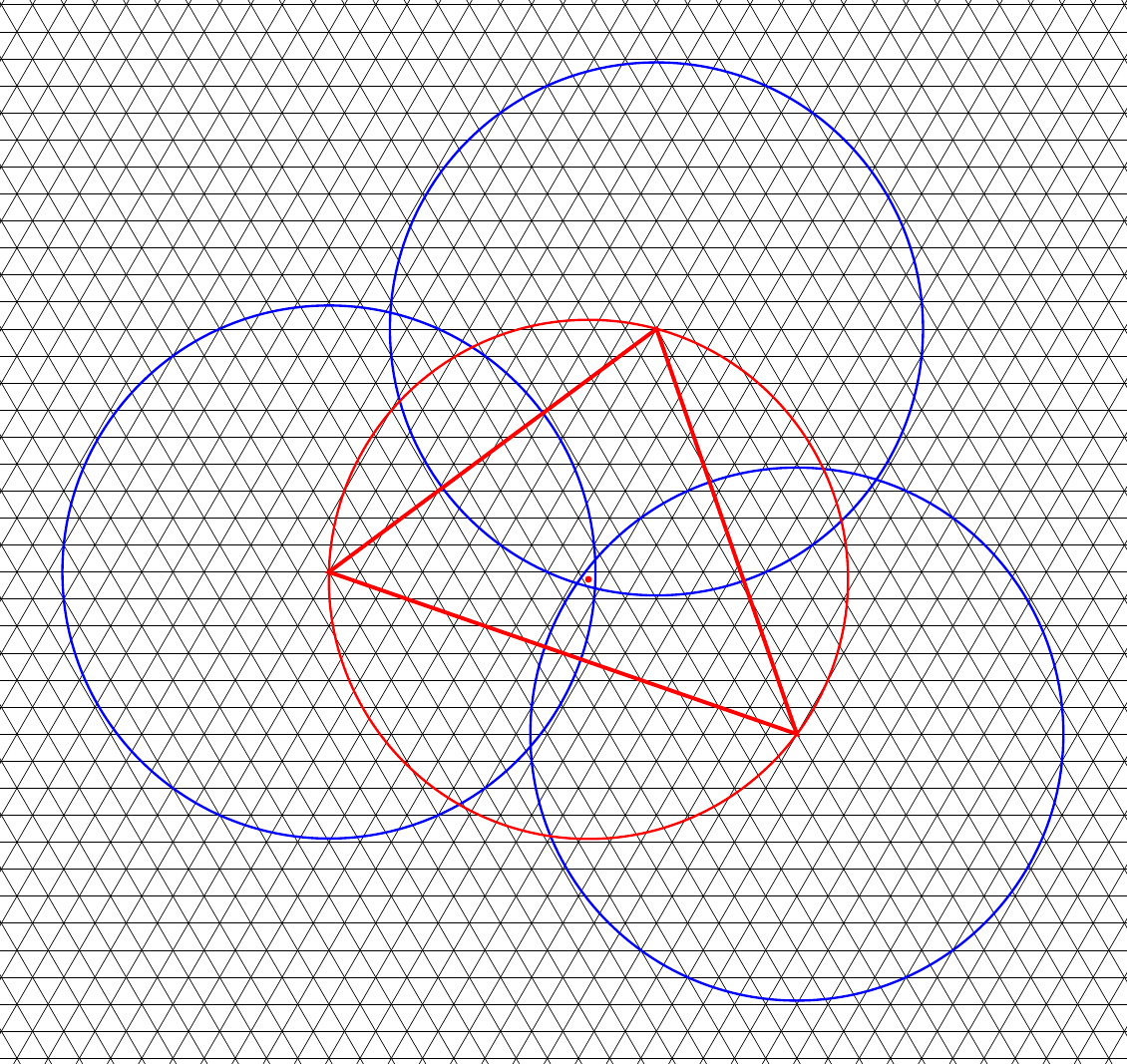}
\caption{A triangle with the squared circumradius $\le\e$. The circumcenter (the red dot) belongs to the intersection of disks of squared radius $\e$ centered at triangle vertices, and the triangle belongs to the union of disks.} 
\end{center}
\end{figure}

Given a L\"oschian number $t$, consider all pairs $m, n\in \N$ with $N(m,n) \le \e$ and set 
$$d(m,n):=N(n-m,n+2m+1)=3N(m,n)+3(m+n)+1. \eqno (6) $$
Let 
$$d_*(\e):=\max\;\Big[d(m,n):\;N(m,n) \le \e,\;d(n,n)< (\sqrt{3\e\,}+1)^2\,\Big]. \eqno(7)$$ Since the maximum in (7) 
is over a finite range, the set of maximizing pairs is non-empty. Consequently, the value $d_*(\e)$
is well-defined and is important because of the following lemma.

\blemc\label{Lemma 3.} 
Given $t=N(m,n)$, every $\sqrt{d_*(\e)}$-equilateral triangle 
$T$ with vertices in $\A_2$ is strongly $\e$-admissible. Such a triangle is unique up to $\A_2$-shifts 
and $\A_2$-symmetries.
\elemc

\begin{proof1} Given $t$ and $d$, the solution to the system
$$\begin{cases}m^2+n^2+mn =t \cr 3t+3(m+n)+1 = d\cr m \le n
\end{cases}$$
is
$$\begin{array}{c}m=\Big[(d-3 t - 1) -\sqrt{36t-3(d-3 t - 1)^2}\Big]\Big/6,\\ 
n=\Big[(d-3 t - 1) +\sqrt{36t-3(d-3 t - 1)^2}\Big]\Big/6,\end{array}$$
and $m=n$ only if $d=(\sqrt{3t}+1)^2$. Thus, the $\sqrt{d}$-equilateral $\A_2$-triangle $T$ is unique up to the lattice shifts and symmetries.

First, we consider the case $0 \le m < n$. For the six combinations of $m,n$ with $0 \le m <n\le 3$ 
the assertion of Lemma 3 is verified directly. For the remaining 
$m,n$ we assume, without loss of generality, that the $\A_2$-coordinates of the vertices 
of $T$ are
$$\bv_0=(m,n),\quad \bv_1=(-m-n-1,m),\quad \bv_2=(n,-m-n-1).$$

Accordingly, $\bo=(-1/3,-1/3)$ is the circumcenter of $T$. Let $\bx_0 \in T$ be 
the $\R^2$-point for which $|\bv_0-\bx_0|^2=|\bv_1-\bx_0|^2=t$ and let $\bx_1$, $\bx_2$ be defined similarly by the cyclic permutation of indexes $0,1,2$. Then
$$r:= |\bo-\bx_i|^2=\dfrac{1}{4}
\left(\sqrt{t + m + n + {1\over3}}-\sqrt{t-3(m + n) - 1} \right)^2=\dfrac{1}{4}\left(\sqrt{{d\over3}}-\sqrt{4t-d} \right)^2.$$

Given $t$ and $s=m+n$, the function $r(m,n)$ achieves its maximum at 
$m=n=s/2$. Furthermore, $r(m,m)$ is monotone decreasing in $m$,  implying that $r(m,n)\le r(3,3)=(31-10\sqrt{6})/3$. Observe that $T \setminus \big(B_t(\bv_1)\cup B_t(\bv_2)\cup B_t(\bv_3)\big) \subset B_r(\bo)$, see Figure~3. 

\begin{figure}[H]
\begin{center}
\includegraphics[scale=0.7]{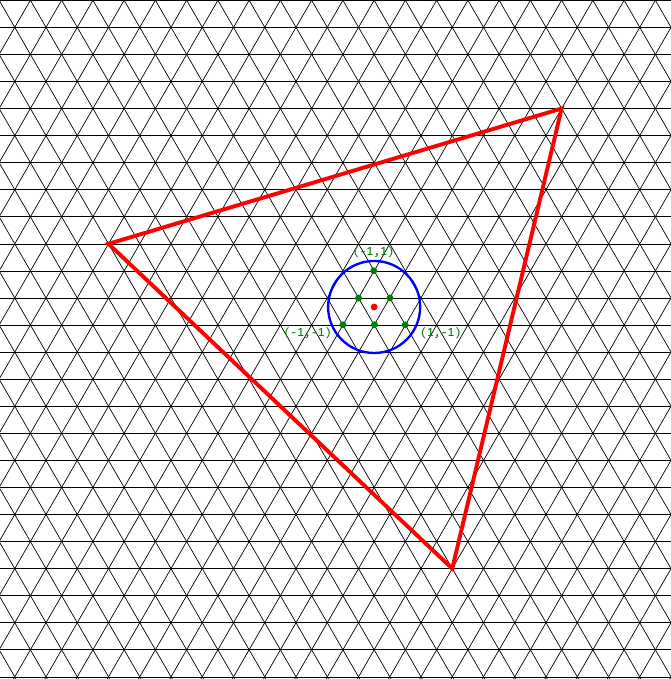}
\caption{A triangle $T$ with the corresponding disk $B_r(\bo)$.} 
\end{center}
\end{figure} 

\noindent
Therefore, we only need to verify that 
$$\min\;\Big[|\bv_i-\bx|^2:\;\bx\in B_r(\bo) \cap \A_2,\,i=1,2,3\Big]\le t.$$ Due to the bound $r <(31-10\sqrt{6})/3$, disk $B_r(\bo)$ contains at most six $\A_2$-sites. It is a direct calculation to verify that for each of them the 
squared distance to the nearest vertex of $T$ is at least $t$. Note that for 
$(1, -1),\, (-1, 1),\, (-1, -1) \in B_r(\bo)$ this minimal squared distance is equal to
$$(m+1)^2 +(n-1)^2 +(m+1)(n-1) = (m^2+n^2+ mn) + 1 + m -n.$$
This quantity does not exceed $t$ because $n-m \ge 1$.

Observe that the case $m=n$ and consequently $\e=3n^2$ and $d=9n^2+6n+1$ is excluded by the condition $d(m,n) < 9n^2+6n+1$, which explains the necessity of this condition.
\end{proof1}

\section{Strongly $\e$-maximal triangles}\label{str_t-max_triangles}

In this section, we analyze triangles of maximal area within the class of strongly $\e$-admissible triangles. Such triangles will be referred to as {\it strongly $\e$-maximal} triangles.

\blemd\label{Lemma 4.}
Given $\e$, the set of strongly $\e$-maximal triangles consists 
of the $\sqrt{d_*(\e)}$-equilateral $\A_2$-triangles only.
\elemd

\begin{proof1} 
Note that the circumcenter of a strongly $\e$-maximal triangle $T$ cannot 
belong to $\A_2$. In fact, if it belongs to $\A_2$, then the squared circumradius of $T$ is at 
most $t$, while there exists a strongly $\e$-admissible triangle from 
Lemma~3 with a larger 
area. Indeed, take a representation $\e=m^2+n^2 + mn$. Then the circumradius of 
$\sqrt{d(m,n)}$-equilateral triangle is $>\sqrt{t}$. Consequently, the area of the 
$\sqrt{d(m,n)}$-equilateral triangle is larger than the area of $T$.

If $T$ is not acute, then by Lemma~1 its area is at most $\left(\sqrt{t}+ 1/\sqrt{3}\right)^2$, which is the area of the isosceles right triangle with circumradius 
$\sqrt{t}+1/\sqrt{3}$. This is smaller than the area ${\sqrt{3}}(3t+3(m+n)+1)/4$ 
of any strongly $\e$-admissible $\sqrt{d(m,n)}$-equilateral triangle from Lemma~3. Thus, $T$ 
is acute. 

Denote by $\bv_i$, $i=0,1,2$, the vertices of $T$, by $\by_{i,j}$ the midpoints of edges $\bv_i\bv_j$, and by $\bo$ the circumcenter of~$T$. Let $W \subset T$ be the unit $\A_2$-triangle containing $\bo$ in its closure. If each vertex of $W$ belongs to a different quadrilateral $Q_i:=\bv_i \by_{i,i+1} \bo  \by_{i-1,i}$, then we have located the desired $W$. (Here $i\pm1$ are understood modulo $3$.) In the opposite case two vertices of $W$ belong to, say $Q_0$, and one belongs to, say $Q_1$ (see Figure~4).

\begin{figure}[h]
\begin{center}
\includegraphics[scale=0.4]{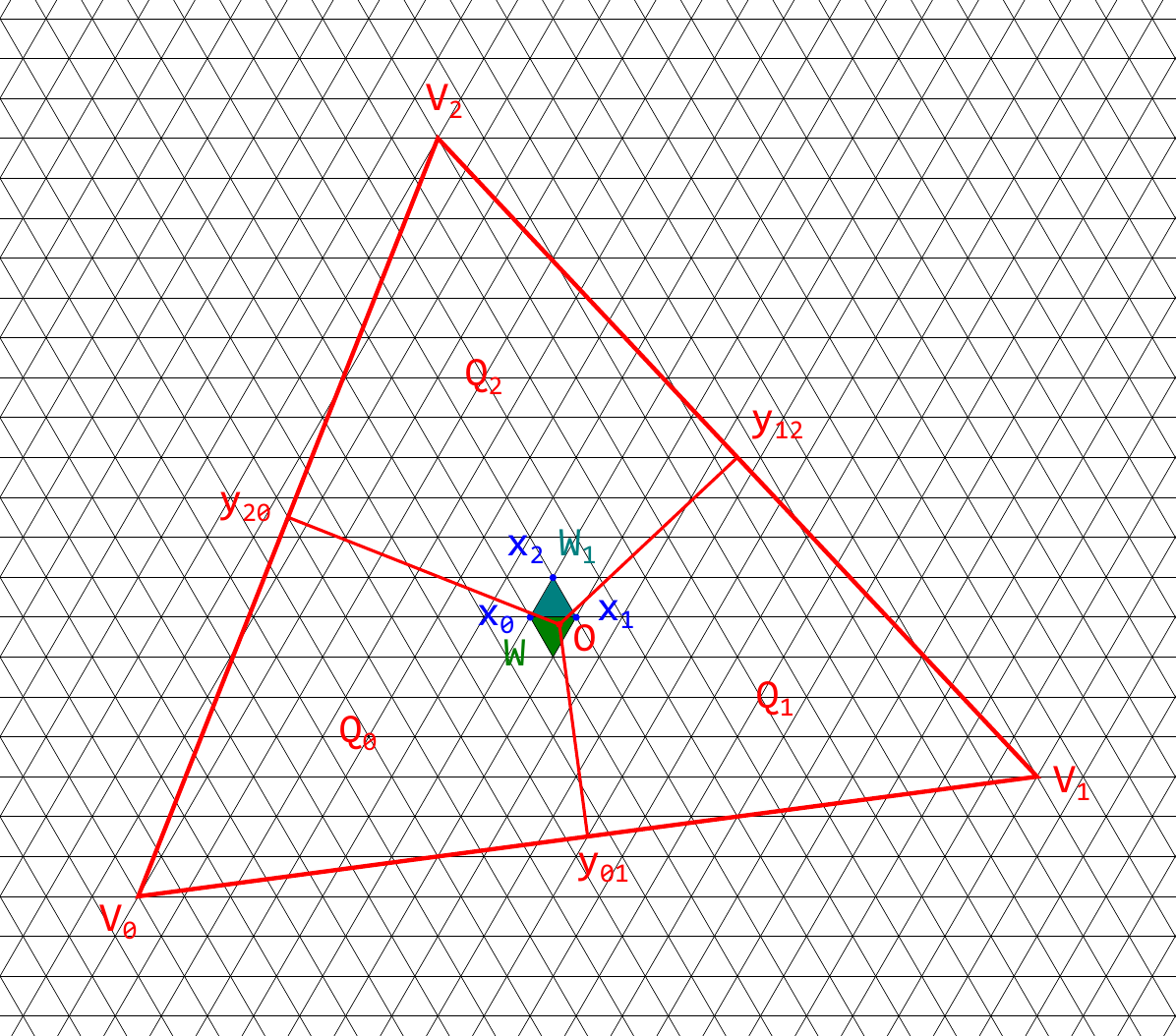}
\caption{Triangles $T$, $W$ and $W_1$.} 
\end{center}
\end{figure}

Consequently, one of the edges of $W$ intersects $Q_2$ but has endpoints 
in $Q_0$ and $Q_1$. Let $W_1$ be the reflection of $W$ about this edge. Then either three vertices of $W_1$ belong to three different $Q_i$ or $W_1$ also has an edge which intersects $Q_2$ but has endpoints in $Q_0$ and $Q_1$. In the later case the reflection process can be continued to produce $W_3$ and so on. As a result, one obtains a broken line consisting of unit edges at the angle $\pi /3$ between two consecutive edges such that each unit edge intersects $Q_2$ but has endpoints in $Q_0$ and $Q_1$. It is clear that this broken line is finite, and the last reflection produces  a unit $\A_2$-triangle $W_k$ with the vertices in three different $Q_i$'s. Indeed, the angle $\angle \by_{0,2}\bo\by_{1,2}$ cannot be contained within a strip of width $\sqrt{3}/2$. Without loss of generality re-denote by $W$ the obtained unit $\A_2$-triangle $W_k$ with the vertices in three different $Q_i$'s.

Denote by $\bx_i$ the vertices of $W$ and by $\bw$ the circumcenter of $W$. Set $C:=\cup_{i=0,1,2}B_t(\bx_i)$. Strong $\e$-admissibility of $T$ implies that the vertices of $T$ belong to $C$. Set $L:=C \setminus B_{d_*(t)/3}(\bw)$, where $d_*(t)$ is defined in (7).  
By construction, $L$ is a union of three lunes $L_i=B_t(\bx_i)\setminus B_{d_*(t)/3}(\bw)$, $i=0, 1, 2$. Recall that the disks $B_{r}(\cdot)$ are closed. Therefore, $L_i\cap B_t(\bx_i)\subset L_i$ while $L_i \cap B_{d_*(t)/3}(\bw) = \emptyset$. Without loss of generality, assume that $\bx_0=(0,0)$,  $\bx_1=(0,-1)$,  
$\bx_2=(-1,0)$, $\bw=(-1/3,-1/3)$ as in the proof of Lemma~3. By construction, the interior of $L$ does not contain $\A_2$-sites. Otherwise, if $(m,n) \in L_0 \setminus \partial L_0$, then $|(m,n)-\bx_0|^2 = N(m,n) < t$ while $|(m,n)-\bw|^2 >  d_*(t)$ which contradicts (7). %(3 vii). 
Thus, the vertices of $T$ (which are $\A_2$-sites) belong to $B_{d_*(t)/3}(\bw) \cap C \subset B_{d_*(t)/3}(\bw)$. Consequently, the circumradius of $T$ does not exceed $d_*(t)/3$ and therefore its area does not exceed the area of a $\sqrt{d_*(t)}$-equilateral $\A_2$-triangle. The latter triangle is strictly $\e$-admissible according to Lemma~3. Hence, $T$ must also be a $\sqrt{d_*(t)}$-equilateral $\A_2$-triangle.
\end{proof1}

\brdd\label{Remark 4.} 
{\rm Observe that, for given $\e$ and the corresponding $d_*(\e)$, the number of 
$\A_2$-sites in the arc 
$$B_t(\bx_0) \cap \{(m,n):\; 0\le m < n,\, 3N(m,n)+3(m+m)+1 =d_*(t) \}$$
can be more than one. In this case there are two or more $\e$-admissible $\sqrt{d_*(\e)}$-equilateral triangles which cannot be taken to each other by $\A_2$-shifts and $\A_2$-symmetries. The identification of all such triangles (contrary to the identification of all $\sqrt{d_*(\e)}$-equilateral triangles) is a tricky number-theoretical question. It is discussed in detail in the companion paper \cite{[MSS3]}. In particular, we show in \cite{[MSS3]} that, up to 
$\A_2$-shifts and $\A_2$-symmetries, the cases in which there are 1, 2, or 3 triangles $T$ occur 
for infinitely many values of $\e$. It is expected that the maximal number of triangles $T$ is
bounded by an absolute constant independent of $\e$. Nevertheless, there are examples of 
values of $\e$ admitting, up to $\A_2$-shifts and $\A_2$-symmetries, 4, 5, 6, or 7 triangles $T$. The only identified case of seven triangles is $d_*(61847325140556)=185542002663199$. }
\erdd

Lemmas~1-4 suggest that the admissible configuration consisting solely of strongly $\e$-maximal triangles is the one with the minimal density. The caveat is the existence of $\e$-admissible but not strongly $\e$-admissible triangles having area 
$>3\sqrt{3}\e/4$. Nevertheless, such a triangle must have an adjacent triangle with 
a considerably smaller area such that the area of their union is $< 3\sqrt{3}\e/2$.

\bleme\label{Lemma 5.} 
For $\e \ge 27$, the area of the union of a $\e$-admissible but not strongly $\e$-admissible triangle with at least one of adjacent Delaunay triangles does not exceed 
$$\sqrt{\e}\left(2\sqrt{\e}+ \frac{1}{\sqrt{3}}+ \sqrt{\left(\sqrt{\e}+  {\frac{1}{\sqrt{3}}}\right)^2-t}\right) 
<\frac{3\sqrt{3}\e}{2}.$$
\eleme

\begin{proof1} 
We need to consider a $\e$-admissible triangle $T=\triangle \bv_0\bv_1\bv_2$ that is not strongly $\e$-admissible and has area larger than $3\sqrt{3}\e/2$. According to Lemmas 1 and 2, the squared circumradius $r$ of $T$ satisfies estimates $\e < r \le (\sqrt{\e}+ {1/\sqrt{3}})^2$. Also, $T \setminus \big (\cup_{i=0,1,2} B_{\e}(\bv_i)\big)$ contains at least one $\bz \in \A_2$. Accordingly, $\bz$ is at squared distance at most $\e$ from some other occupied site $\bv_3$ which is the vertex of a Delaunay triangle adjacent to $T$. Without loss of generality assume that $\bv_1\bv_2$ is the common edge of these two triangles, and $|\bv_1-\bv_2|=2a$. 
 
If $\sqrt{\e} \le a \le  \sqrt{\e}+  {1/\sqrt{3}}$, then the distance between $\bv_1\bv_2$ and the circumcenter $\bo$ of $T$ is at most $\sqrt{(\sqrt{\e}+  {1/\sqrt{3}})^2-a^2}$ such that the distance from $\bv_0$ to $\bv_1\bv_2$ is at most $\sqrt{\e} + 1/\sqrt{3}+ \sqrt{(\sqrt{\e}+  {1/\sqrt{3}})^2-a^2}$. Simultaneously, the distance from $\bv_3$ to $\bv_1\bv_2$ is at most $\sqrt{\e}$ as it is shorter than $|\bv_3-\bz|$. Hence, the area of the union of two triangles is at most
$$a\left(2\sqrt{\e}+ 1/\sqrt{3}+ \sqrt{(\sqrt{\e}+ {1/\sqrt{3}})^2-a^2}\right).$$

\begin{figure}[h]
\begin{center}
\includegraphics[scale=0.34]{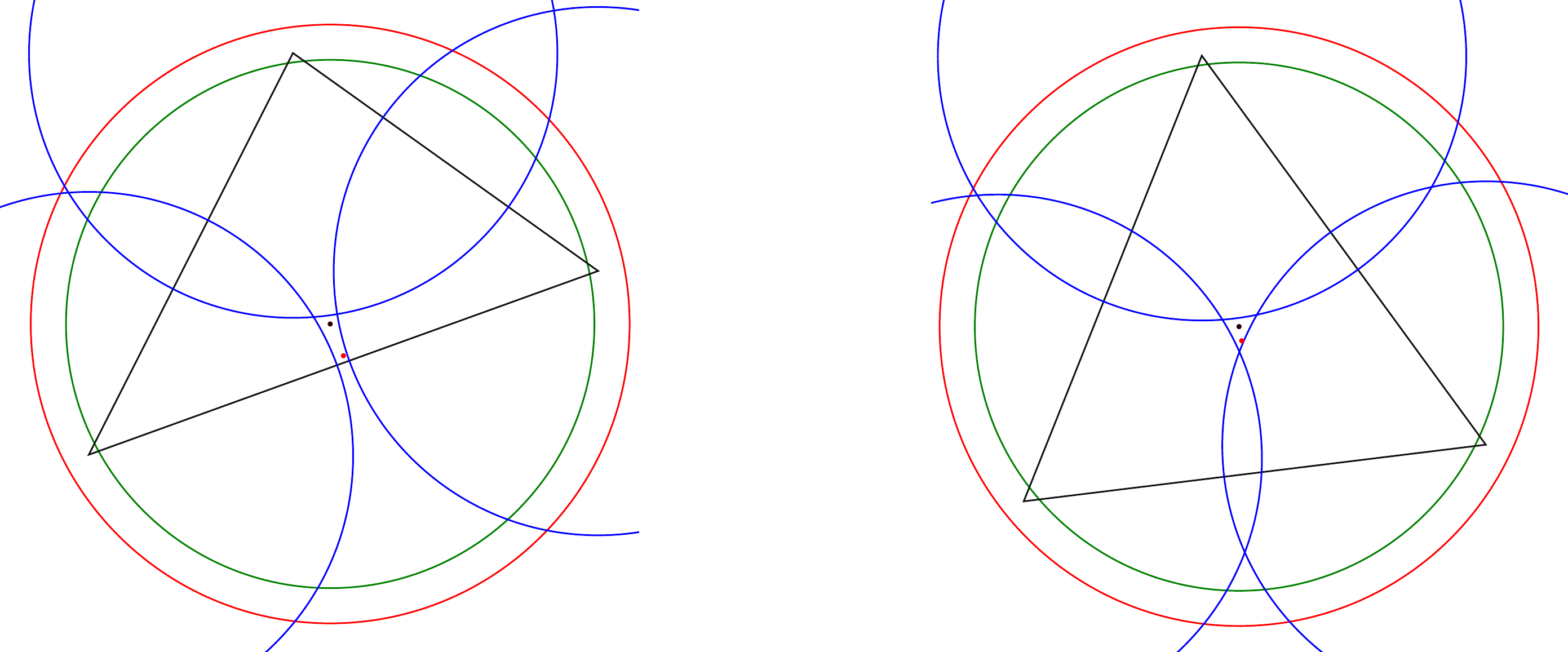}
\caption{Cases $\sqrt{\e} \le a \le  \sqrt{\e}+  {1/\sqrt{3}}$ (left) and $1 \le a \le  \sqrt{\e}$ (right). The circumcenter of~$T$ with vertices in $\A_2$ is colored black, while point $\bz\in \A_2$ is colored red. The green and red circles are of radii $\sqrt{\e}$ and $\sqrt{\e}+  {1/\sqrt{3}}$, respectively. The blue circles have radius $\sqrt{\e}$ and are centered at the vertices of $T$.} 
\end{center}
\end{figure}

If $1 \le a \le  \sqrt{\e}$, then the distance from $\bz$ to $\bv_1\bv_2$ is at least $\sqrt{t-a^2}$ and therefore the distance from $\bv_3$ to $\bv_1\bv_2$ is at most $\sqrt{\e}-\sqrt{t-a^2}$. Hence, the area of the union of two triangles is at most
$$a\left(2\sqrt{\e}+ 1/\sqrt{3} + \sqrt{(\sqrt{\e} +  {1/\sqrt{3}})^2-a^2}-\sqrt{t-a^2}\right).$$
A direct calculation shows that, for a given $\e$, the maximum of the above functions of $a$ is attained at the common point $a=\sqrt{\e}$ and is equal to
$$\sqrt{\e}\left(2\sqrt{\e}+ 1/\sqrt{3}+ \sqrt{(\sqrt{\e}+  {1/\sqrt{3}})^2-t}\right).$$
(See Figure~5.) For $\e \ge 27$ this quantity is smaller than ${3\sqrt{3}\over2}\e$.
\end{proof1}

Hence, the $\e$-admissible but not strongly $\e$-admissible triangles are irrelevant while searching for the Delaunay triangles with maximal area. To disregard them we re-partition the corresponding quadrilateral $\bv_0\bv_1\bv_3\bv_2$ along the diagonal $\bv_0\bv_3$ (instead of the canonical partitioning along $\bv_1\bv_2$). This results in two triangles $\triangle \bv_0\bv_1\bv_3$ and $\triangle \bv_0\bv_2\bv_3$. By construction, each of them has the area smaller than the maximal area ${3\sqrt{3}\over4}\e$ of a strongly $\e$-admissible triangle.

A similar re-partitioning works if a given $\e$-admissible 
but not strongly $\e$-admissible triangle requires two or three occupied neighbors $\bv_3,\bv_4$ or $\bv_3,\bv_4,\bv_5$ to produce an occupied site at the distance at most $\e$ from every site inside $T \setminus \big (\cup_{i=0,1,2} B_{\e}(\bv_i)\big)$. The union of these triangles is a pentagon or a hexagon, and we again re-partition them via different diagonals. The specific triangulations used in previous sections are exactly these re-partitioned Delaunay triangulations. Their characteristic property is that every constituting triangle in such a re-partitioned triangulation has area at most ${3\sqrt{3} \over 4}\e$.

From the perspective of Theorems 1 -- 3 the cases where $\e < 27$ can be investigated by a direct enumeration, and they lead to the same conclusions
as for $\e \ge 27$.

\bigskip
\noindent
{\bf Acknowledgments}

\medskip
\noindent
IS and YS thank Math Department, PSU, for support. YS is grateful to DPMMS, University of Cambridge, and St John's College, Cambridge, for support.

\end{document}